\documentclass[11pt]{article}
\usepackage{blois,epsfig}

\def\Journal#1#2#3#4{{#1} {\bf #2}, #3 (#4)}

\def\NIMA{{\em Nucl. Instrum. Methods} A}
\def\NPA{{\em Nucl. Phys.} A}

\def\PRL{\em Phys. Rev. Lett.}

\def\PRC{{\em Phys. Rev.} C}

\graphicspath{{figures/}}

\def\be{\begin{equation}}
\def\ee{\end{equation}}
\def\bea{\begin{eqnarray}}
\def\eea{\end{eqnarray}}

\begin{document}
\vspace*{4cm}
\title{RESULTS OF THE NEMO-3 DOUBLE BETA DECAY EXPERIMENT}

\author{ M. BONGRAND\\ for the NEMO-3 Collaboration}

\address{LAL, Universit\'e Paris-Sud 11, CNRS/IN2P3, Orsay, France.}

\maketitle\abstracts{
The NEMO-3 experiment is searching for neutrinoless double beta decay ($0\nu\beta\beta$) for 2 main isotopes ($^{100}$Mo and $^{82}$Se) and is studying the two-neutrino double beta decay ($2\nu\beta\beta$) of seven isotopes. The experiment has been taking data since 2003 and, up to the end of 2009, showed no evidence for neutrinoless double beta decay. Two 90\% CL lower limits on the half-lives of the transitions were obtained : $\mathcal{T}_{1/2}^{0\nu} > 1.0~10^{24}$~yr for $^{100}$Mo and $\mathcal{T}_{1/2}^{0\nu} > 3.2~10^{23}$~yr for $^{82}$Se. The corresponding limits on the effective Majorana neutrino mass are respectively $\langle m_{\nu} \rangle < 0.47 - 0.96$~eV and $\langle m_{\nu} \rangle < 0.94 - 2.5$~eV. The measurements of the two-neutrino double beta decays for all the isotopes have also reached the highest precision to date.}

\section{Introduction}
\label{sec:introduction}

Experimental search for the neutrinoless double beta decay ($0\nu\beta\beta$) is of major importance in particle physics because if observed it will reveal the Majorana nature of the neutrino ($\nu \equiv \overline{\nu}$) and may allow an access to the absolute neutrino mass scale. The decay violates the lepton number and is therefore a direct probe for the physics beyond the standard model. The existence of this process may be related to right-handed currents in electroweak interactions, supersymmetric particles with R-parity nonconservation, and massless Goldstone bosons, such as majorons.

In the case of the neutrino-mass mechanism the $0\nu\beta\beta$ decay (Fig.~\ref{fig:graphs}) rate can be written as
{\small{$$[\mathcal{T}_{1/2}^{0\nu}(A,Z)]^{-1} = G_{0\nu}~(\mathcal{Q}_{\beta\beta},Z)~|\mathcal{M}_{0\nu}(A,Z)|^2~\langle m_{\nu} \rangle^2$$}
where $\langle m_{\nu} \rangle$ is the effective neutrino mass, $\mathcal{M}_{0\nu}$ is the nuclear matrix element (NME) and $G_{0\nu}$ is the kinematical factor proportional to the transition energy to the fifth power, $\mathcal{Q}_{\beta\beta}^5$.

The spontaneous two-neutrino double beta decay ($2\nu\beta\beta$) is a rare second-order weak interaction process. The accurate measurement of the $2\nu\beta\beta$ decay is important since it constitutes the ultimate background in the search for the 0ν decay signal. It is the testing ground for nuclear models and provides a valuable input for the theoretical calculations of the $0\nu\beta\beta$ decay NME.

The objective of the NEMO-3 experiment is the search for the $0\nu\beta\beta$ decay and study of the $2\nu\beta\beta$ decay with 10 kg of $\beta\beta$ isotopes.

\begin{figure}[htb]
  \begin{center}
    \includegraphics[height=0.11\textheight]{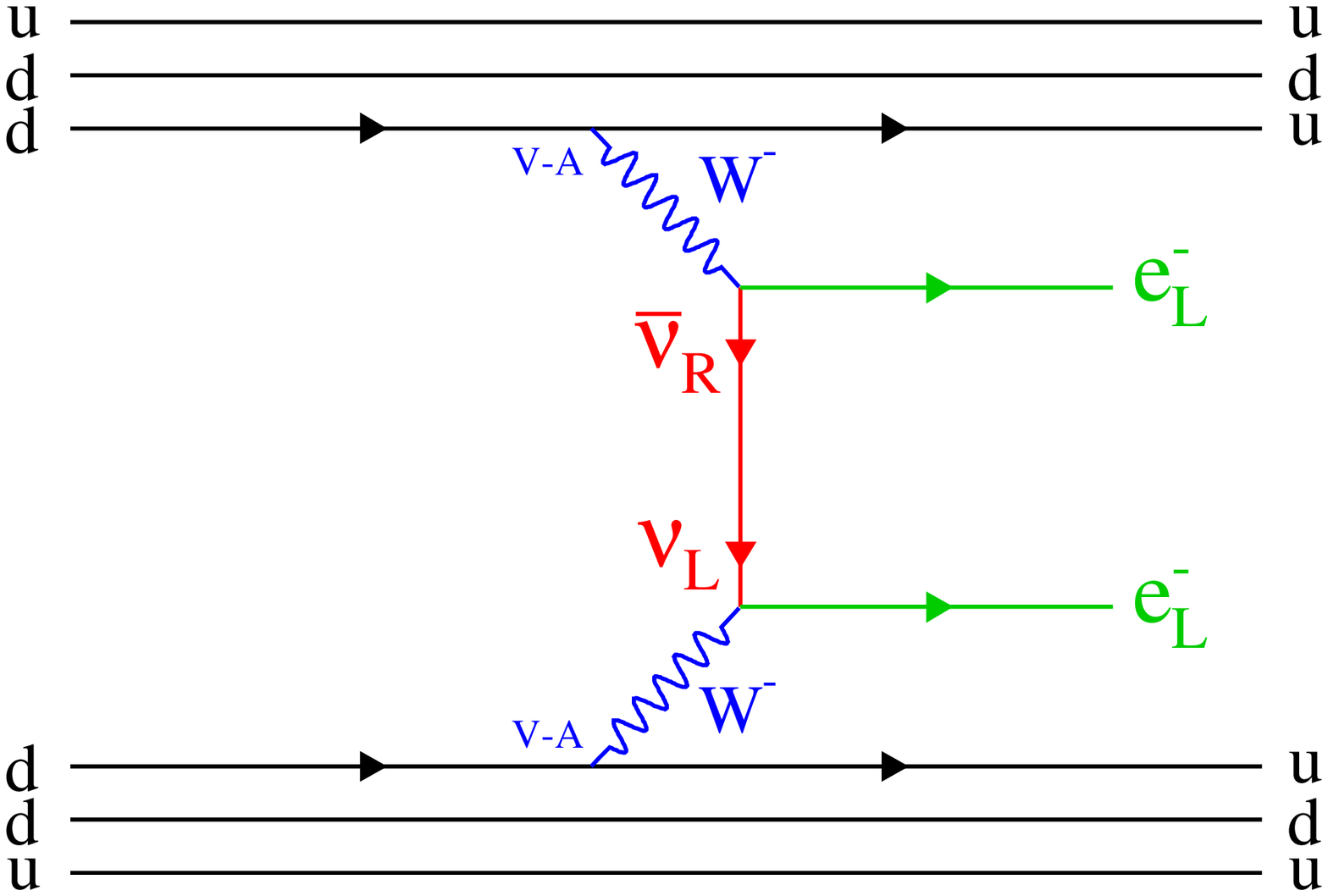}\hspace{5mm}
    \includegraphics[height=0.11\textheight]{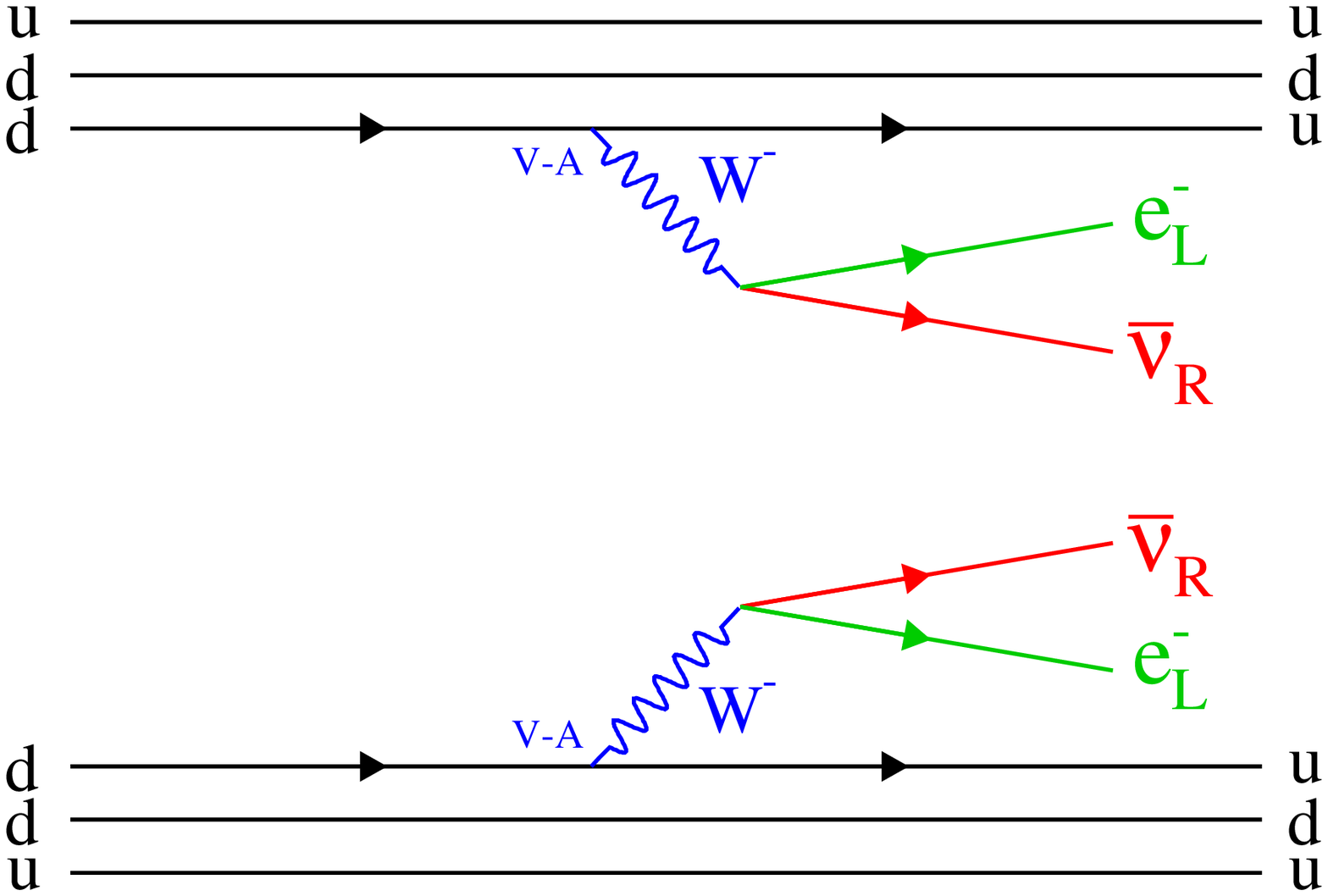}
    \vspace{-2mm}
    \caption{Neutrinoless (light neutrino exchange) and two-neutrino double beta decay mechanisms.}
    \label{fig:graphs}
  \end{center}
\end{figure}

\section{The NEMO-3 experiment}
\label{sec:nemo3}

The NEMO-3 experiment is currently running in the \textit{Laboratoire Souterrain de Modane} (LSM) in the Fr\'ejus tunnel between France and Italy at the depth of 4800 mwe. The design of the NEMO-3 detector~\cite{tdr} is for the direct detection of two electrons from double beta decay by a tracking chamber and a calorimeter measuring individual energies and times-of-flight (Fig.~\ref{fig:nemo3}). Two phases of data have to be considered: high radon and low radon phases, because of the radon-free air facility installation in the LSM in 2004 flushing a tight tent around the detector.
 
\begin{figure}[htb]
  \begin{center}
    \includegraphics[width=0.32\textwidth]{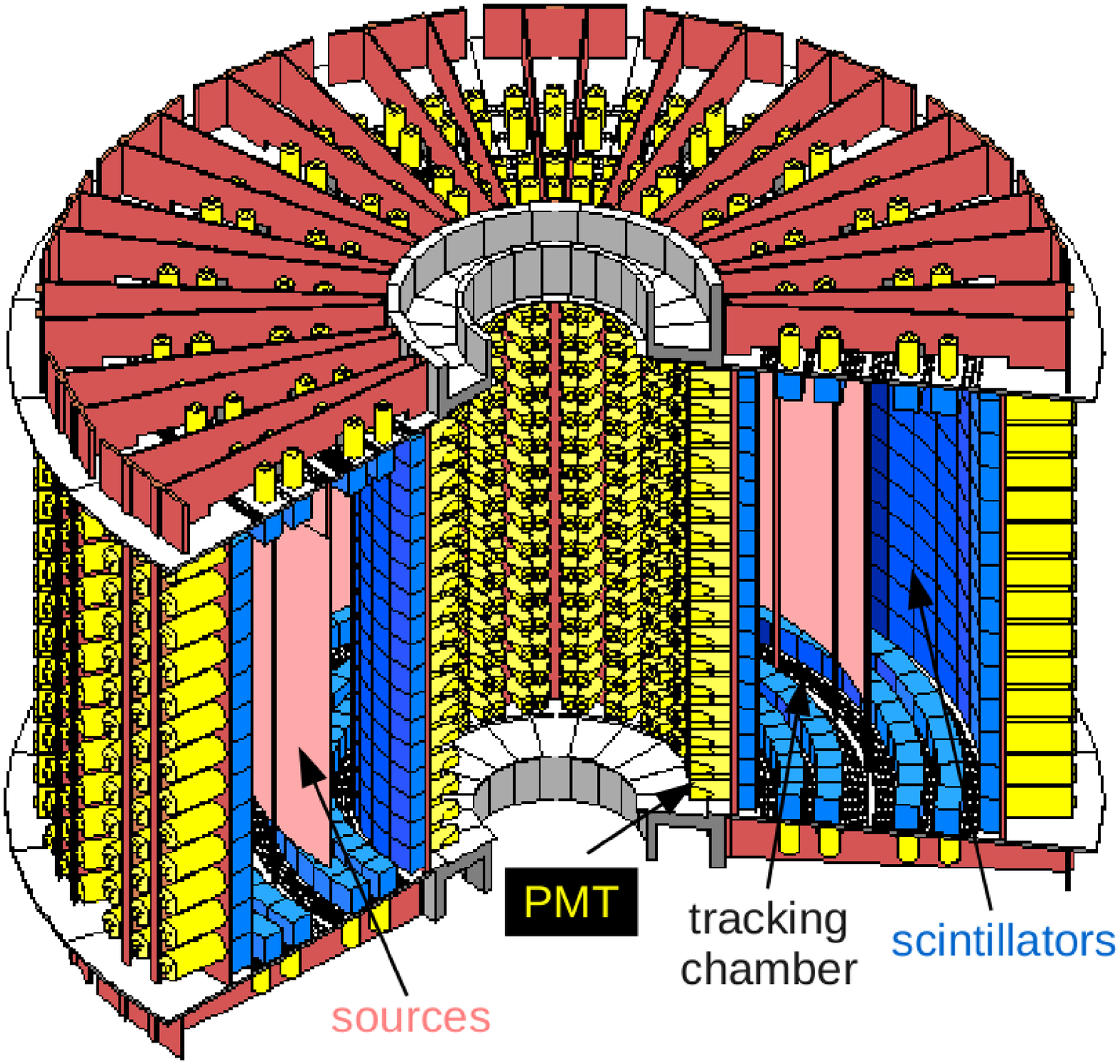}\hspace{2mm}
    \includegraphics[height=0.18\textheight]{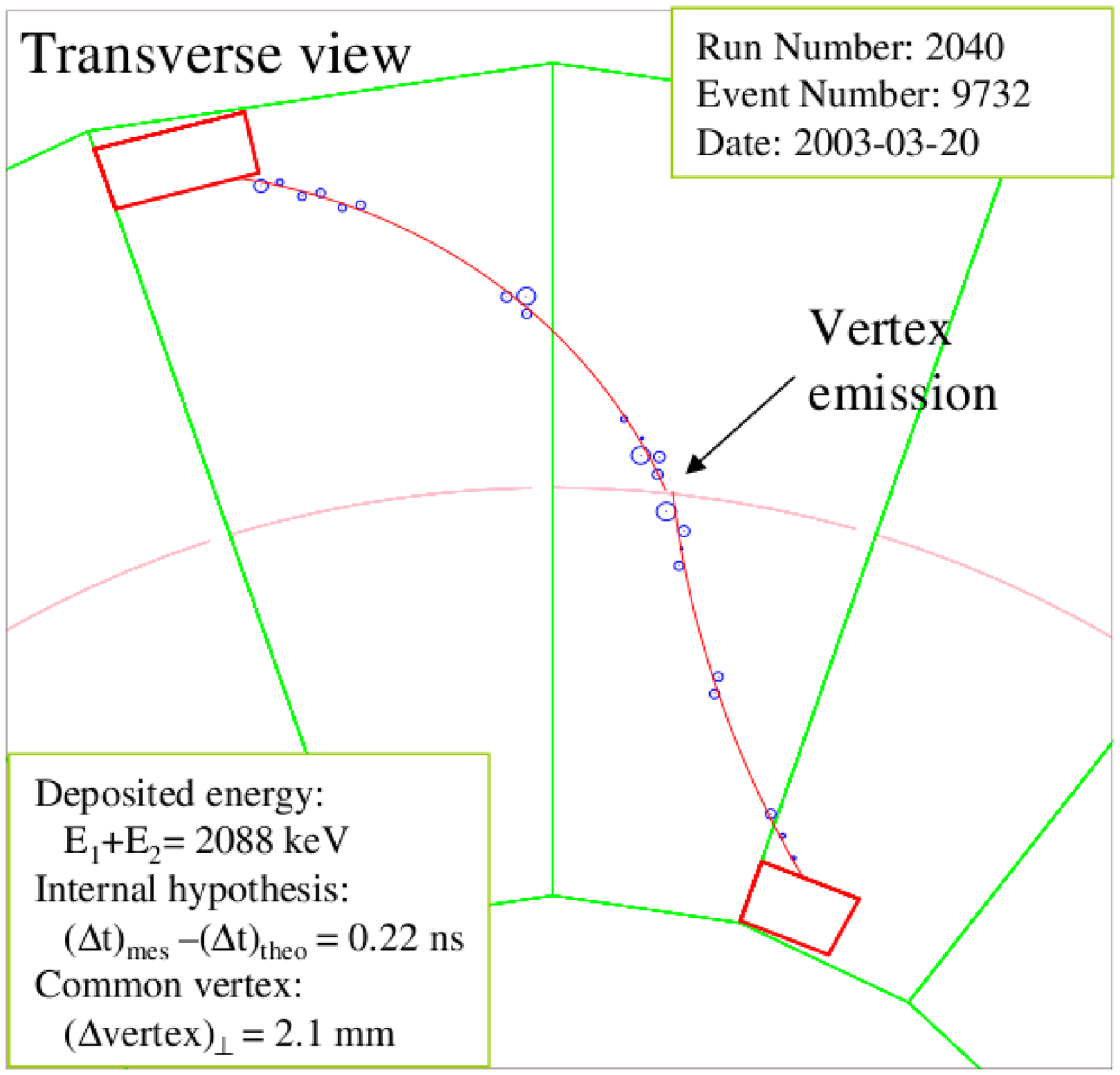}\hspace{2mm}
    \includegraphics[height=0.18\textheight]{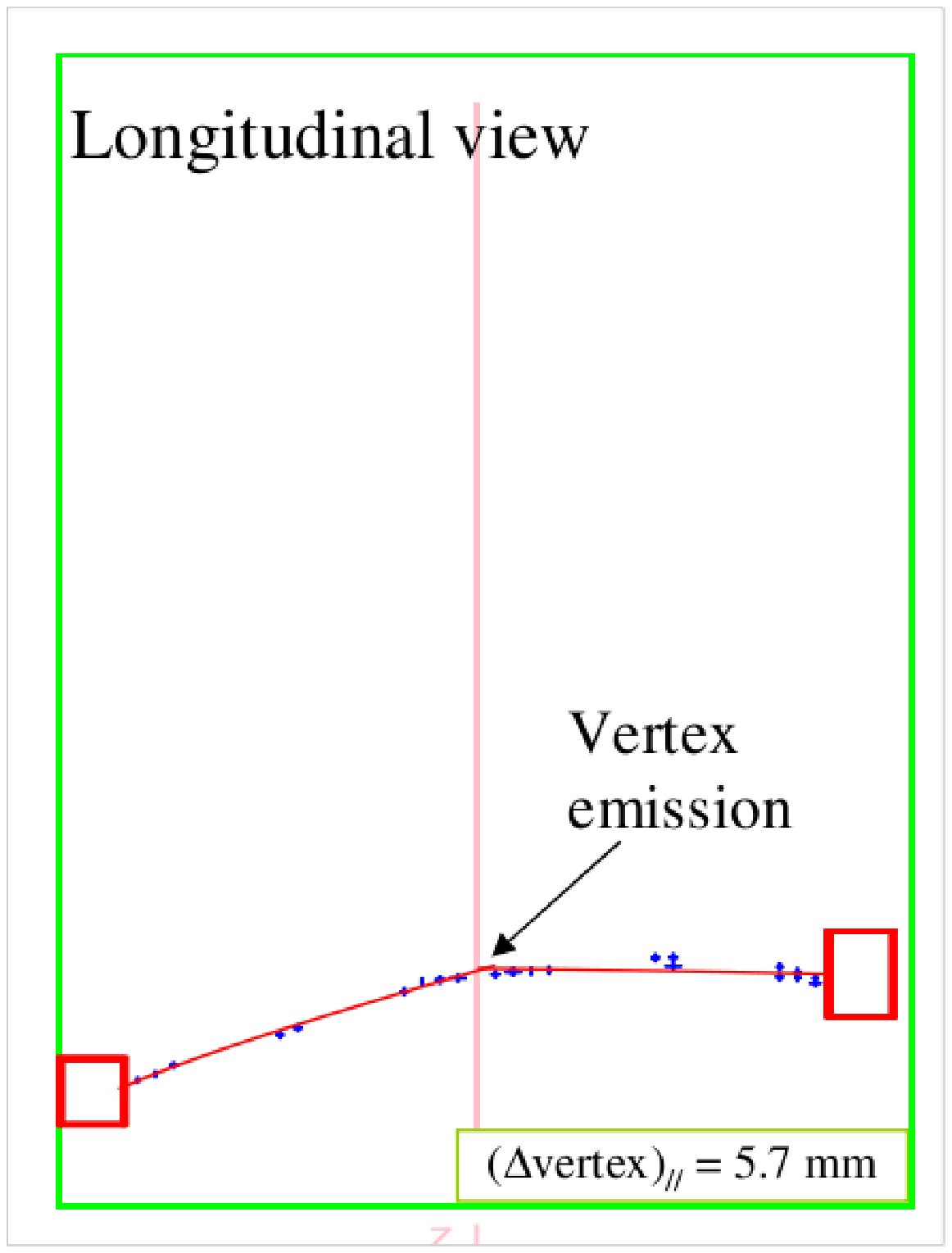}
    \caption{total and individual energies and the angular distribution.}
    \label{fig:nemo3}
  \end{center}
\end{figure}

The detector has a cylindrical shape and is protected by pure iron, wood and borated water shielding. Thin source foils ($\sim$ 50 mg/cm$^{2}$) are located in the middle of the tracking volume surrounded by the calorimeter. Almost 10~kg of enriched isotopes (listed in Table \ref{tab:t12}) were used to produce the source foils. The 6.9~kg of $^{100}$Mo and the 0.9~kg $^{82}$Se are used to search for $0\nu\beta\beta$ decay, smaller amounts of $^{130}$Te, $^{116}$Cd, $^{150}$Nd, $^{96}$Zr and $^{48}$Ca are used to measure two-neutrino double beta decay. The sectors with pure copper, natural and enriched tellurium are used to study the external background. The tracking chamber contains 6180 open drift cells operating in Geiger mode. It provides a vertex resolution of about 1~cm. The calorimeter consists of 1940 plastic scintillator blocks with photomultiplier readout. The energy resolution FWHM is 14-17\%/$\sqrt{E}$. The time resolution of 250 ps allows excellent suppression of the crossing electrons background. A 25 Gauss magnetic field is used for charge identification. The detector is capable of identifying e$^-$, e$^+$, $\gamma$ and $\alpha$ particles and allows a precise measurements of the background components and a good discrimination between signal and background events.

A typical double beta decay candidate is shown in Fig.~\ref{fig:nemo3}. The tracking algorithm finds two electron tracks (with curvatures corresponding to negative charges) with a common vertex in a source foil. Each track ends as it enters a plastic scintillator and must deposit an energy greater than 200~keV. These tracks allow us to determine the angular distribution of the $\beta\beta$ events which, together with the individual energies of the electrons, are important kinematic parameters to study the $\beta\beta$-decay mechanisms. This unique feature is obtained by the combination of the track-calorimetric approach of the NEMO-3 experiment. In order to reject external background events a time-of-flight analysis is also used. 

\section{Backgrounds Measurements}
\label{sec:backgrounds}

As discussed before, the NEMO-3 detector is able to identify different type of particles and combinations. It is possible to define analysis channels to study and measure specifically each background. For example, to measure the external $\gamma$-rays flux entering the inner detector we can look at crossing electrons produced by Compton effect in a scintillator when the electron escapes this one (Fig.~\ref{fig:backgrounds}). Fitting the energy distribution of these events we determine the contributions of the natural radioactivity isotopes contaminating parts of the detector (the glass of the low radioactivity PMTs is still the main contribution).

To measure the radon background inside the tracking chamber of the detector, we use the dedicated electronics to identify the delayed alpha tracks after the $\beta$ decays from the $^{214}$Bi$\rightarrow$$^{214}$Po processes of $^{222}$Rn daughters. The Fig.~\ref{fig:backgrounds} shows a BiPo event close to the source foil and the time distribution in good agreement with the $^{214}$Po half-life.

We can also look at single $\beta$ or $\beta-n\gamma$ decays in the source, external $e^-\gamma$ events...

\begin{figure}[htb]
  \begin{center}
    \fbox{\includegraphics[width=0.24\textwidth]{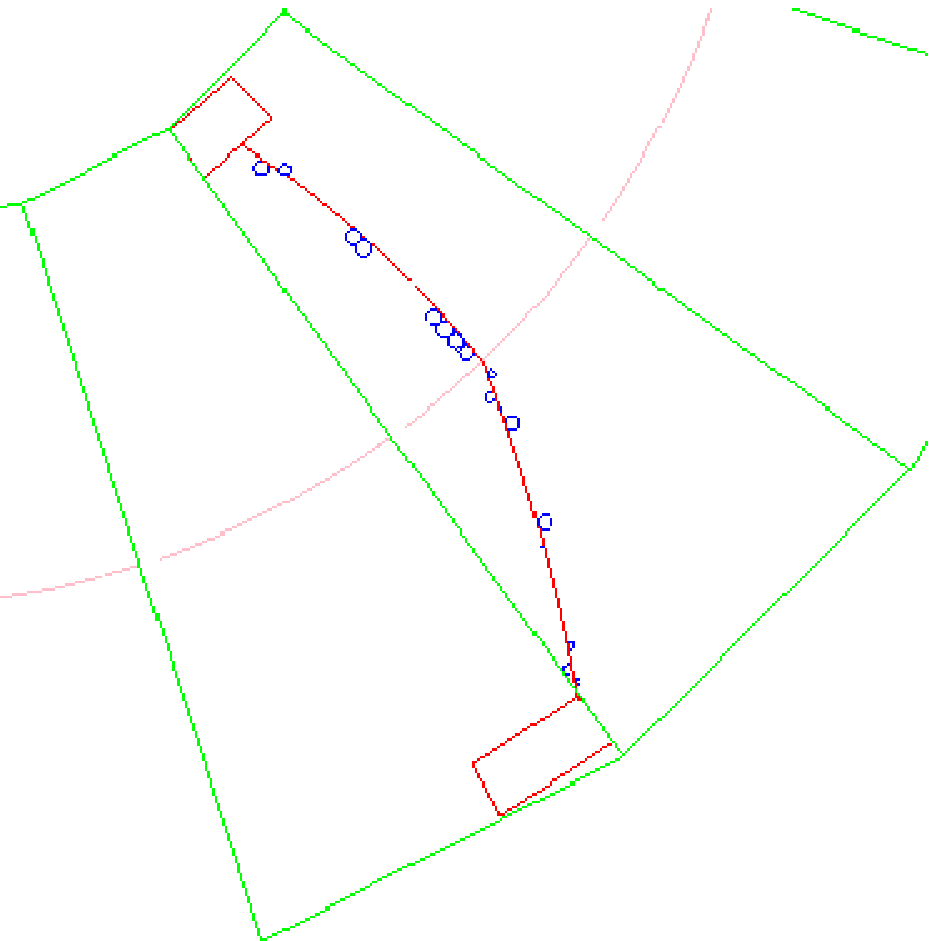}}\hspace{5mm}
    \includegraphics[width=0.31\textwidth]{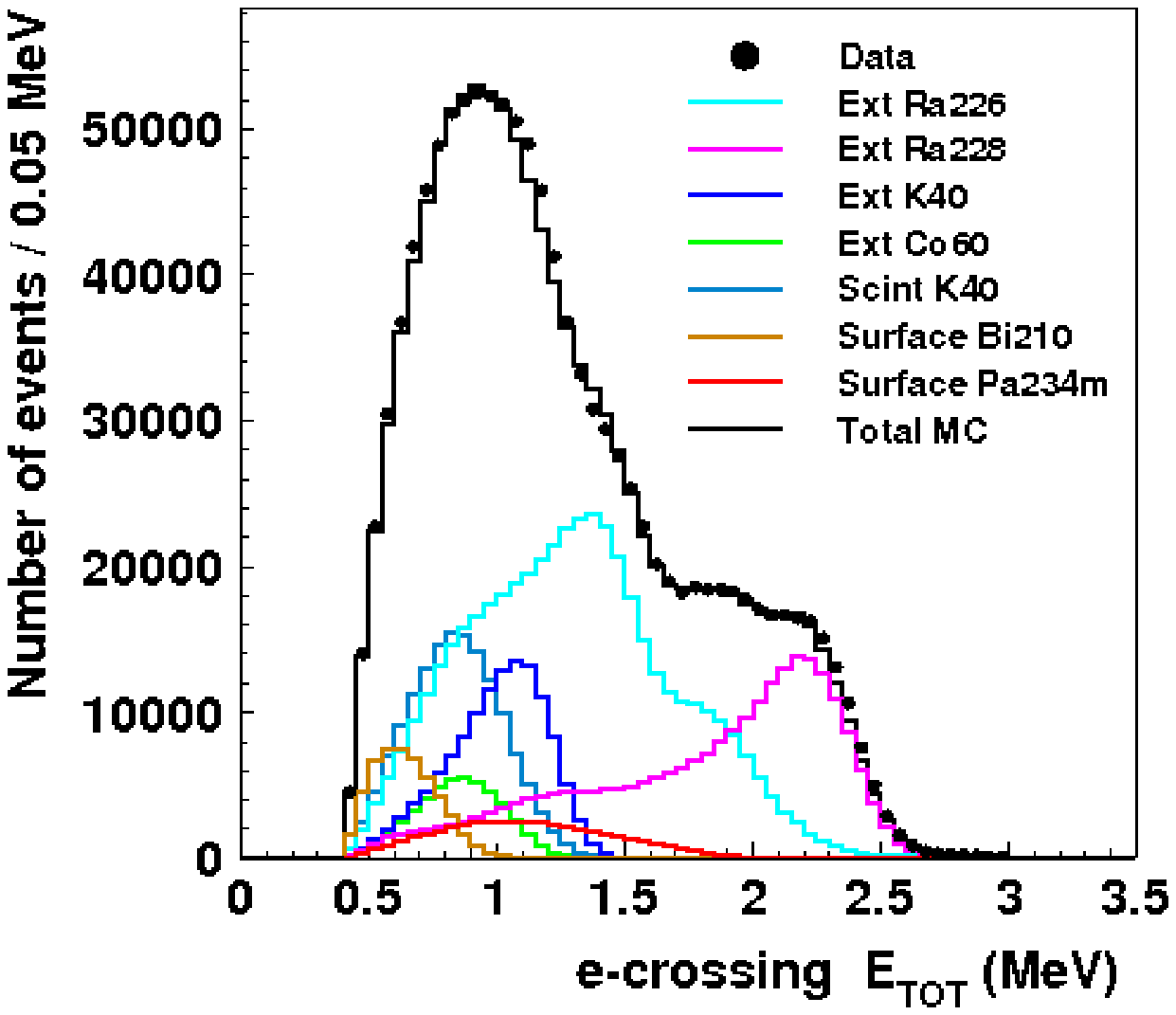}
    \caption{A crossing electron event and the corresponding fit to the data.}
    \label{fig:backgrounds}
  \end{center}
\end{figure}

\begin{figure}[htb]
  \begin{center}
    \fbox{\includegraphics[width=0.24\textwidth]{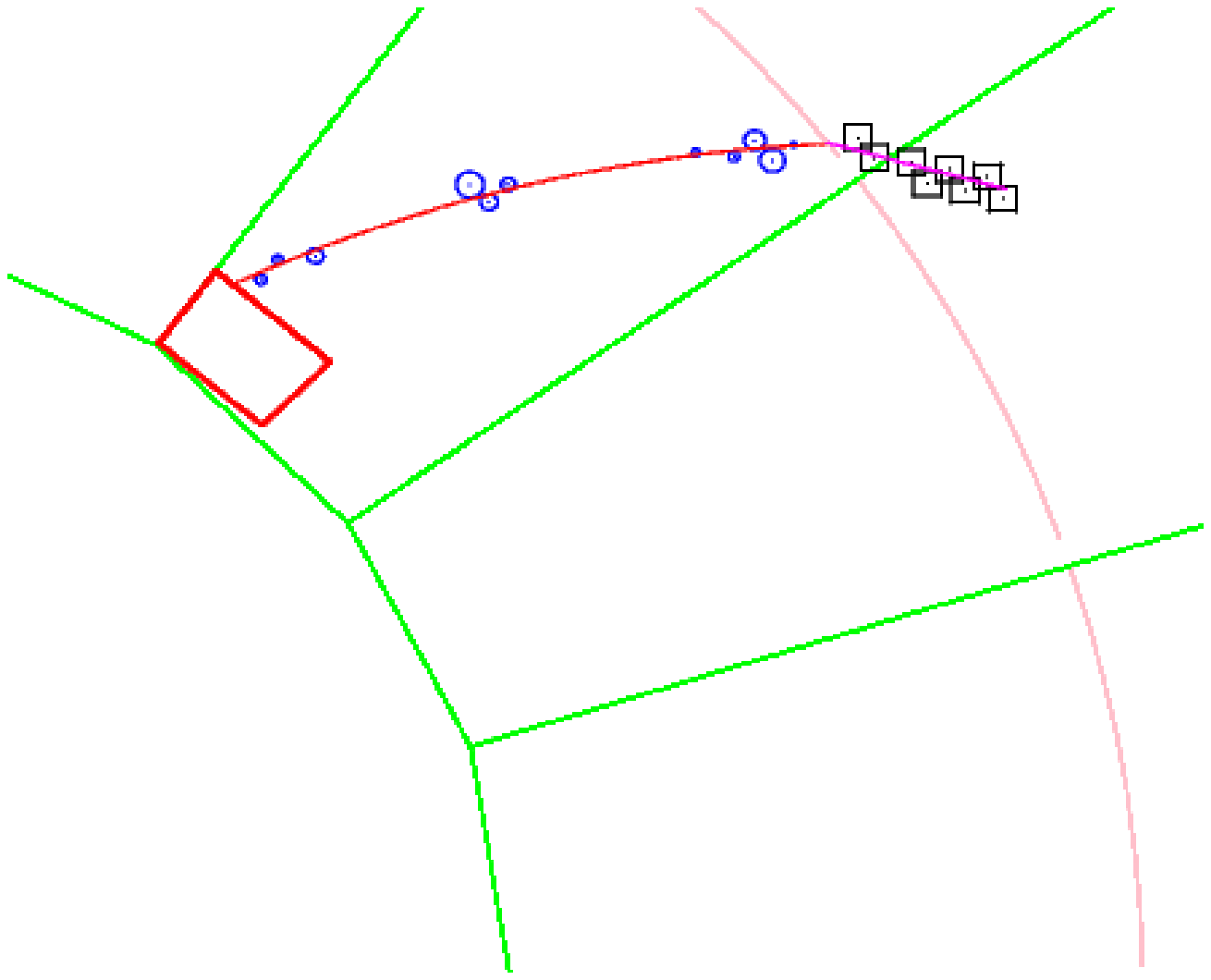}}\hspace{5mm}
    \includegraphics[width=0.33\textwidth]{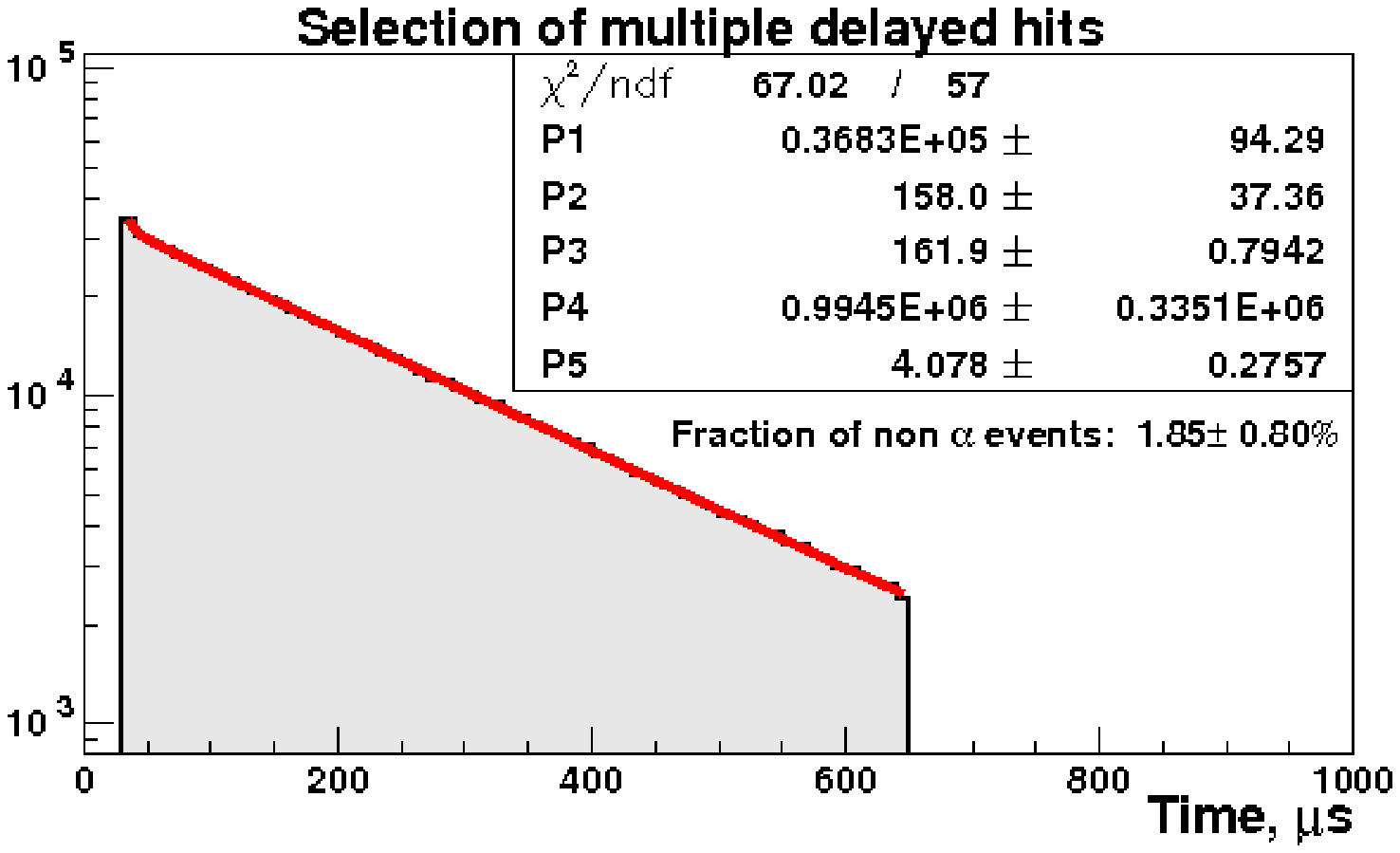}
    \caption{A $^{214}$BiPo process event close to the foil and time distribution in good agreement with the $^{214}$Po half-life of 164~$\mu$s.}
    \label{fig:backgrounds}
  \end{center}
\end{figure}

\section{Two-Neutrino Double Beta Decay Measurements}
\label{sec:dbd}

The measurements of the $2\nu\beta\beta$ decay half-lives were performed for the 7 isotopes of NEMO-3 (see Table~\ref{tab:t12}). The most precise measurement is for the main isotope $^{100}$Mo because of the high mass and the high signal to background ratio due to a low $\mathcal{T}_{1/2}^{2\nu}$ half-life. With this result we can appreciate the accuracy of the understanding of the NEMO-3 data by looking at the distribution of the total and individual energies and the angle between the 2 electrons (Fig.~\ref{fig:100mo}).

\begin{figure}[htb]
  \begin{center}
    \includegraphics[width=0.32\textwidth]{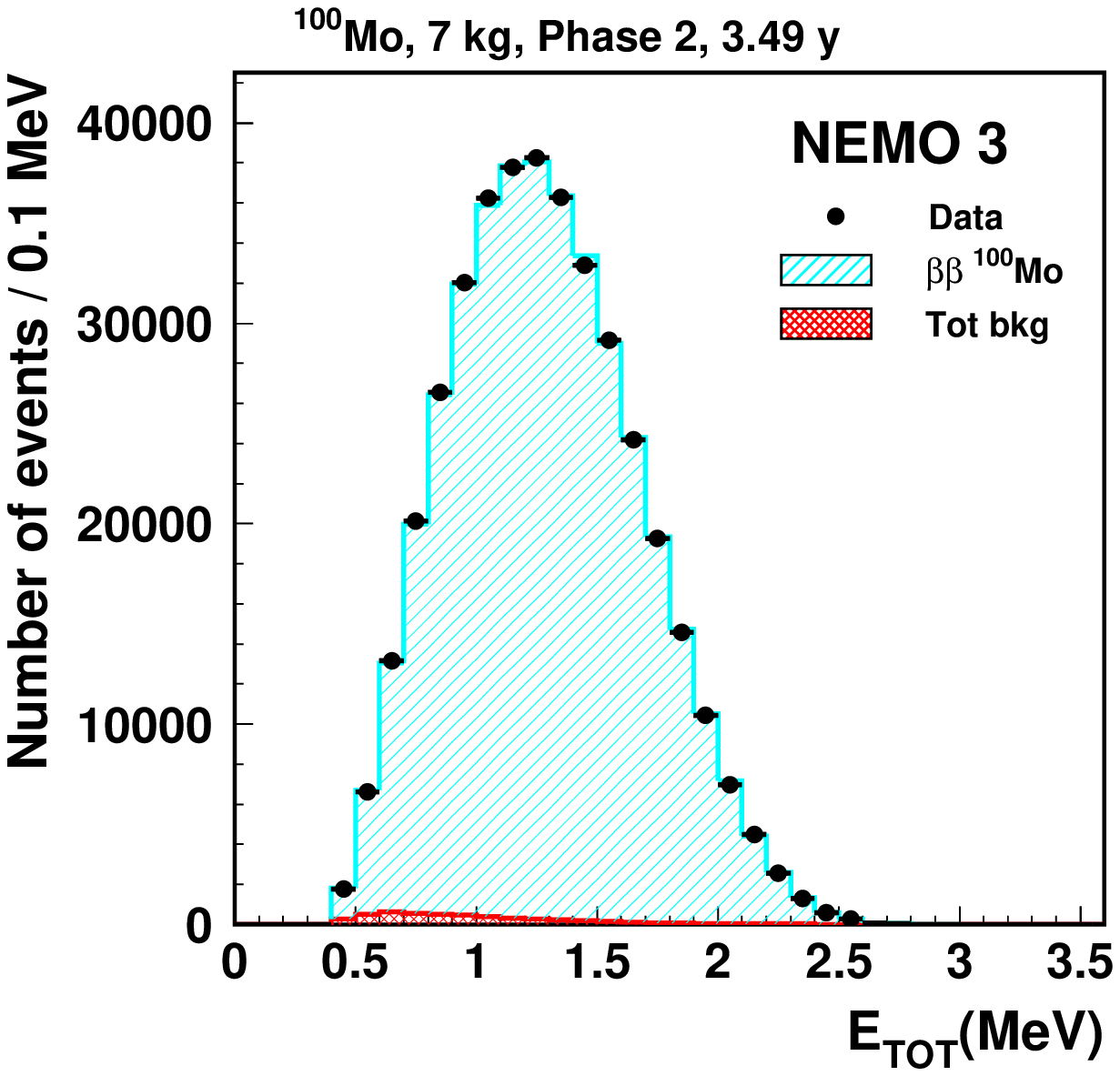}\hspace{1mm}
    \includegraphics[width=0.32\textwidth]{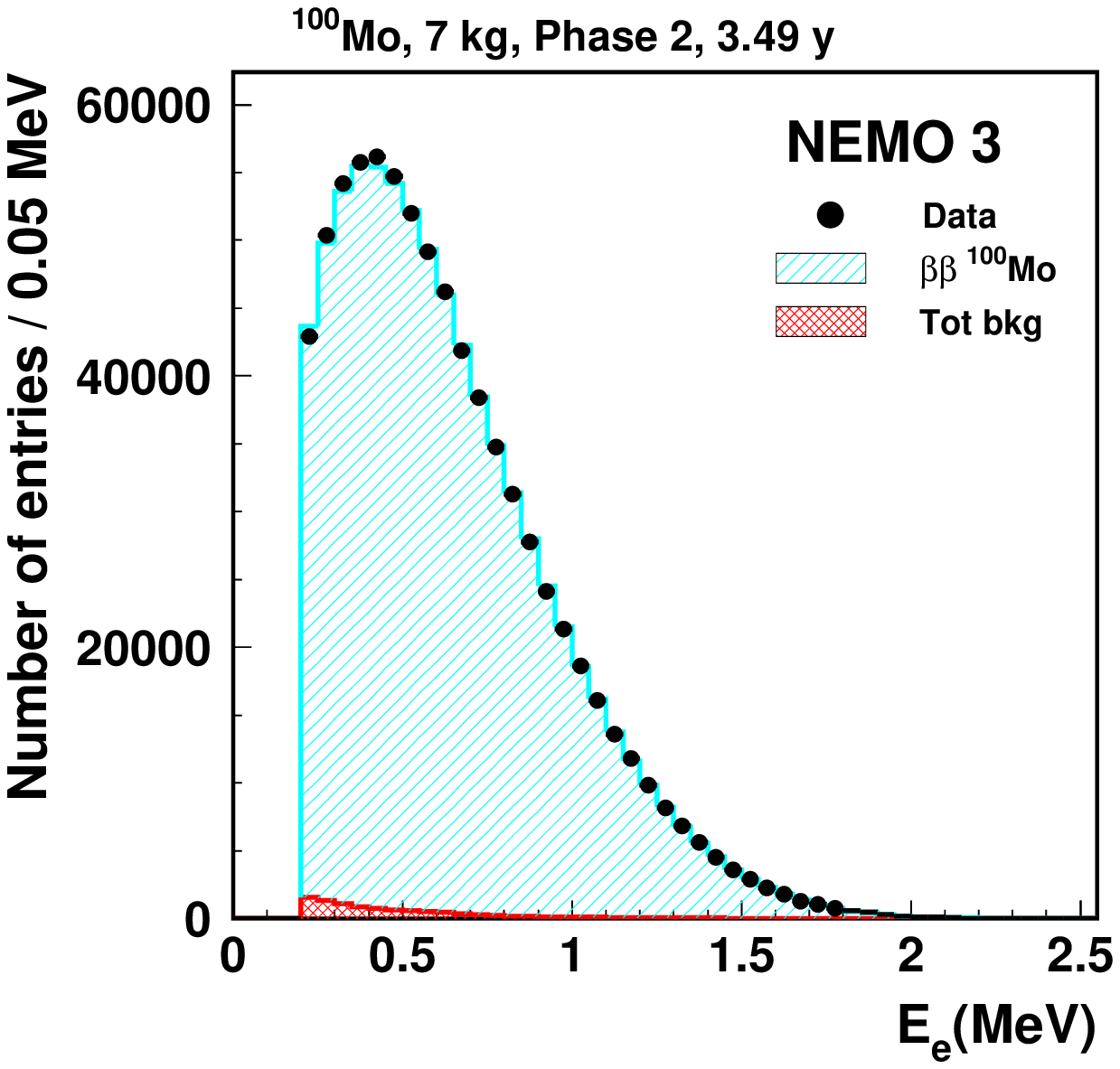}\hspace{1mm}
    \includegraphics[width=0.32\textwidth]{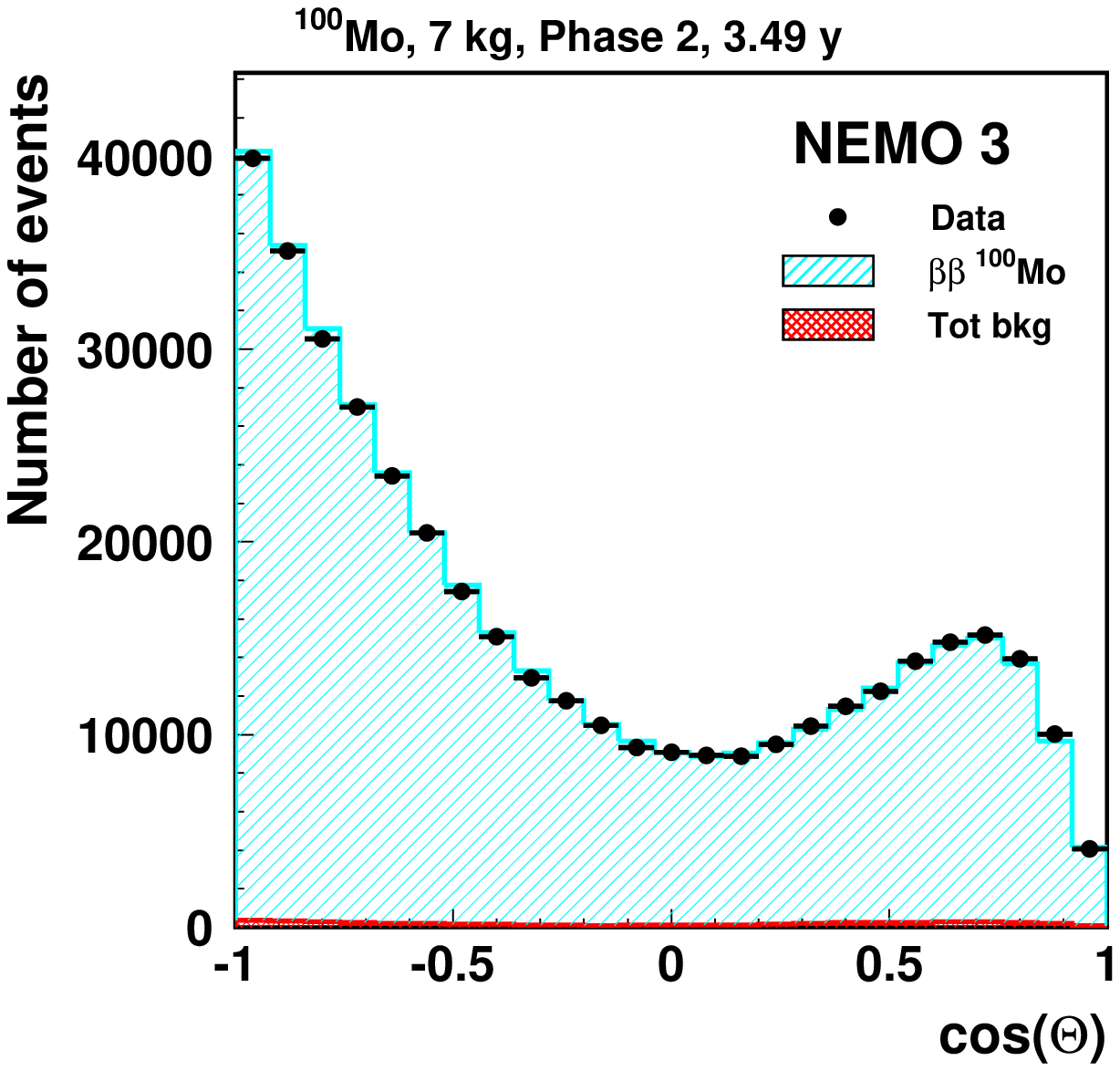}
    \caption{Total energy, individual energy and angular distributions of the $^{100}$Mo $2\nu\beta\beta$ events in the NEMO-3 experiment for the low radon data phase (3.49 years).}
    \label{fig:100mo}
  \end{center}
\end{figure}

From this analysis, the half-life of the $^{100}$Mo was extracted with the nuclear matrix elements for the two-neutrino process, important for the theoretical calculations. Same analyses were performed for all the other isotopes and the results are summarized in Table~\ref{tab:t12}. These results are the most precise direct measurements of $2\nu\beta\beta$ rates today.

\begin{table}[hhh]
  \caption{NEMO~3 results of the $2\nu\beta\beta$ half-life measurements.
    \label{tab:t12}}
  \vspace{3mm}
  \begin{center}
    \small
    \begin{tabular}{ |c|c|c|c|c|c| }
      \hline
      Isotope& Mass [g] & $\mathcal{Q}_{\beta\beta}$ [keV] & Sig/Bkg & $\mathcal{T}_{1/2}$ [$10^{19}$ years] & $\mathcal{M}^{2\nu}$ \\
      \hline
      $^{100}$Mo &6914& 3034 & 76   & 0.717 $\pm$ 0.001 (stat) $\pm$ 0.054 (syst) & 0.126 $\pm$ 0.006 \\
      $^{82}$Se  &932 & 2995 & 4    & 9.6 $\pm$ 0.1 (stat) $\pm$ 1.0 (syst) & 0.049 $\pm$0.004 \\
      $^{130}$Te &454 & 2529 & 0.25 & 70 $^{+10}_{-8}$ (stat) $^{+10}_{-9}$ (syst)& 0.017 $\pm$ 0.003 \\
      $^{116}$Cd & 405& 2805 & 10.3  & 2.88 $\pm$ 0.04 (stat) $\pm$ 0.16 (syst) & 0.069 $\pm$ 0.003 \\
      $^{150}$Nd &37.0& 3368 & 2.8  & 0.920 $\pm$ 0.025 (stat) $\pm$ 0.063 (syst) & 0.030 $\pm$ 0.002 \\
      $^{96}$Zr  &9.4 & 3350 & 1.0  & 2.35 $\pm$ 0.14 (stat) $\pm$ 0.16 (syst) & 0.049 $\pm$ 0.002 \\
      $^{48}$Ca  &6.99 & 4274 & 6.8  & 4.4 $^{+0.5}_{-0.4}$ (stat) $\pm$ 0.4 (syst) & 0.024 $\pm$ 0.002 \\
      \hline
    \end{tabular}
  \end{center}
\end{table}

The two-neutrino double beta decay to excited states has also been studied in NEMO-3, with a measurements for the $0^+ \rightarrow 0^+_1$ transition: $\mathcal{T}_{1/2}^{2\nu} = 5.7~^{+1.3}_{-0.9}~stat.~\pm~0.8~syst.~10^{20}$~yr and a 90 \% CL limit for the $0^+ \rightarrow 2^+_1$ decay: $\mathcal{T}_{1/2}^{2\nu} > 1.1~10^{21}$~yr.

\section{Neutrinoless Double Beta Decay Search}
\label{sec:ndbd}

The neutrinoless double beta decay search was performed on the NEMO-3 data from 2003 to the end of 2009 and no evidence for neutrinoless double beta decay has been observed in $^{100}$Mo nor in $^{82}$Se (Fig.~\ref{fig:0nu}). Therefore, 90\% CL lower limits on the half-lives have been set: $\mathcal{T}_{1/2}^{0\nu} > 1.0~10^{24}$~yr for $^{100}$Mo and $\mathcal{T}_{1/2}^{0\nu} > 3.2~10^{23}$~yr for $^{82}$Se. The corresponding limits on the effective Majorana neutrino mass are respectively $\langle m_{\nu} \rangle < 0.47 - 0.96$~eV and $\langle m_{\nu} \rangle < 0.94 - 2.5$~eV, according to the most recent NME calculations used by NEMO-3 \cite{Kort_mo} \cite{Kort_se} \cite{Simkovic}.

\begin{figure}[htb]
  \begin{center}
    \includegraphics[width=0.35\textwidth]{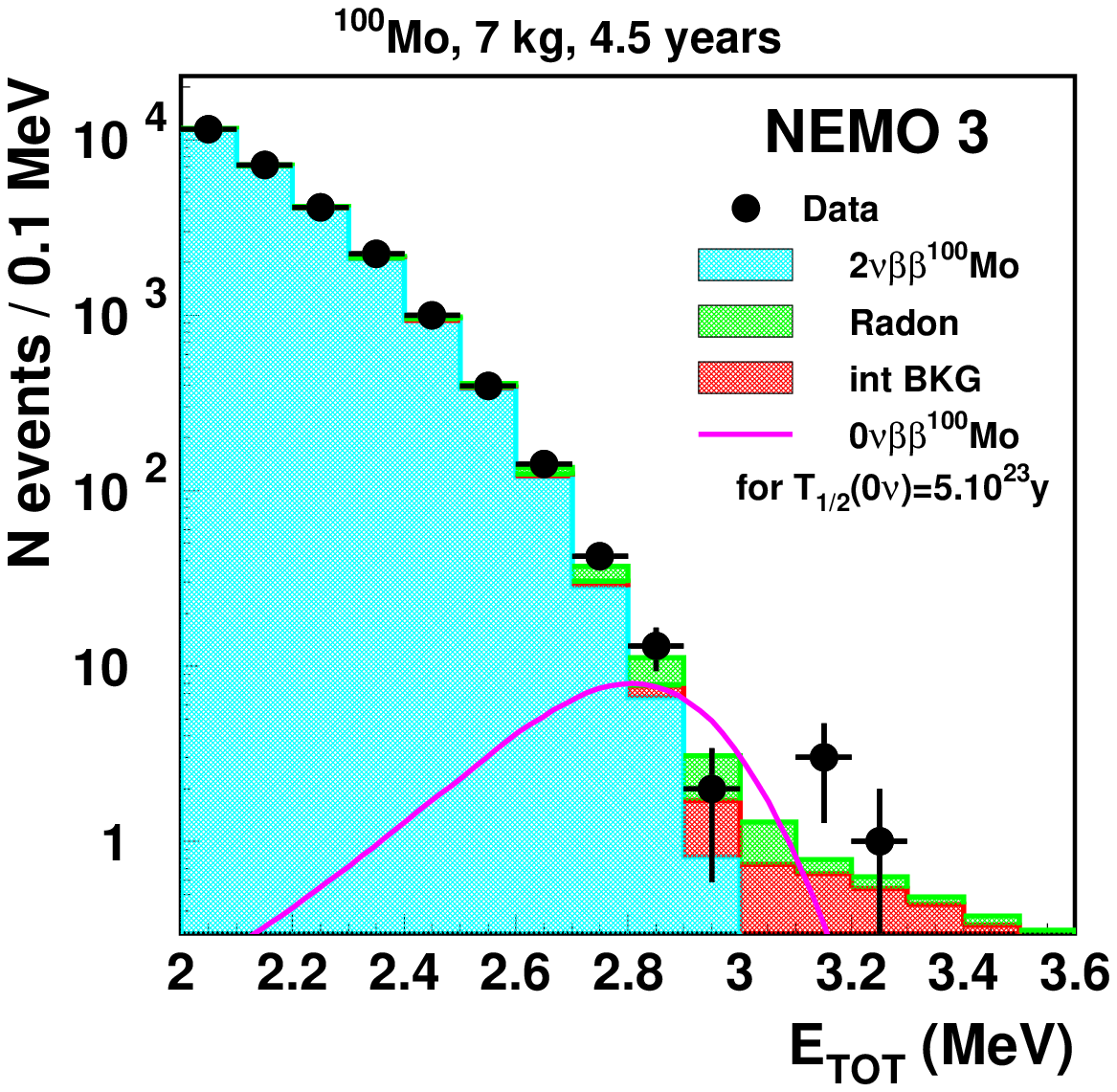}\hspace{1cm}
    \includegraphics[width=0.35\textwidth]{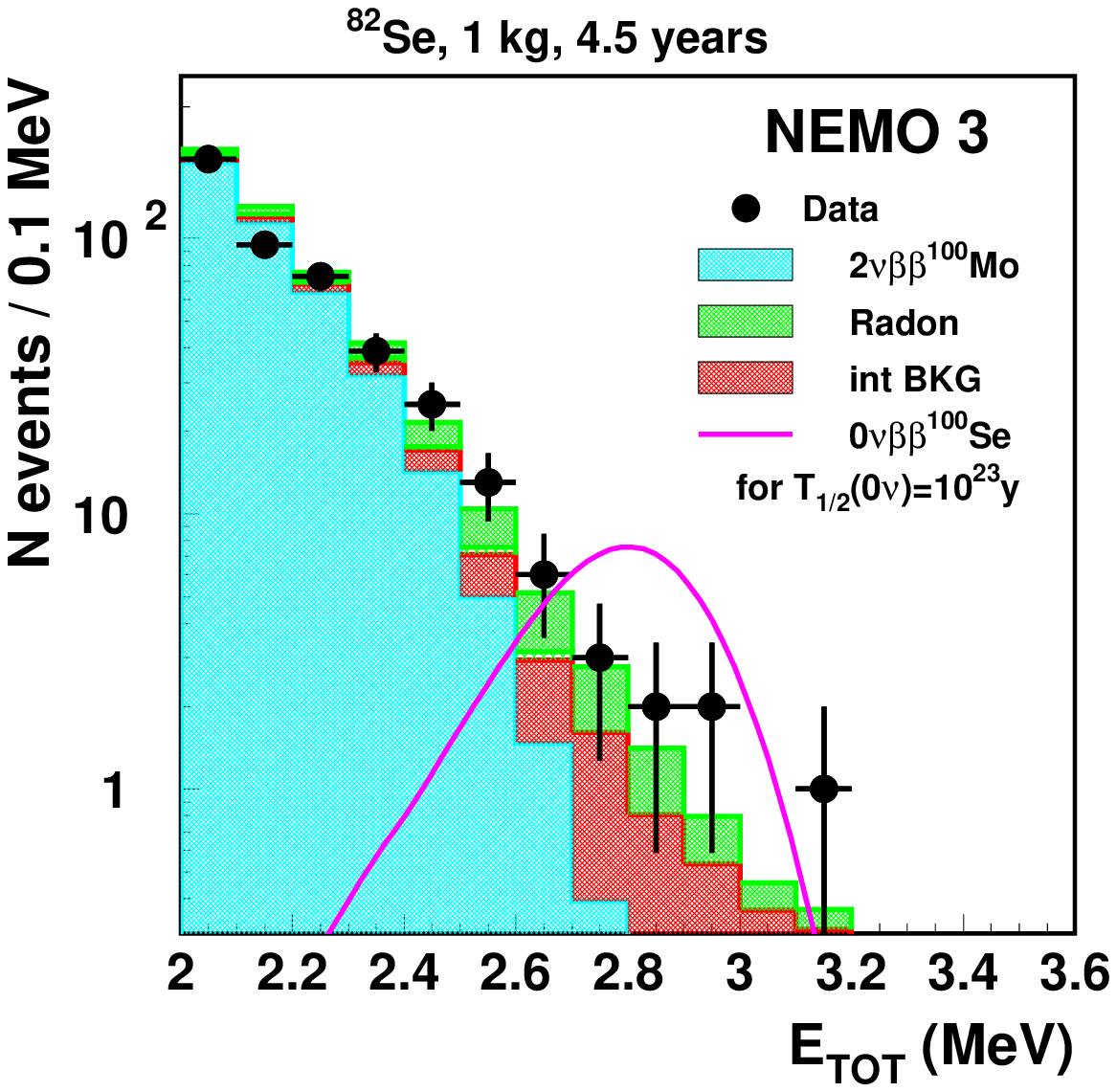}
    \caption{Total energy spectra of 2 electrons events observed in NEMO-3 after 4.5 years for $^{100}$Mo on the left and $^{82}$Se on the right. For $^{100}$Mo, 18 events have been observed between 2.8 and 3.2 MeV for 16.4 $\pm$ 1.4 expected. For $^{82}$Se, 14 events have been observed between 2.6 and 3.2 MeV for 10.9 $\pm$ 1.3 expected. For illustration, the magenta line represents what a $0\nu\beta\beta$ signal would look like with a given half-life.}
    \label{fig:0nu}
  \end{center}
\end{figure}

Other $0\nu\beta\beta$ mechanisms have also been investigated and several limits (90 \% CL) have been set for decays to excited states: $\mathcal{T}_{1/2}^{0\nu}(0^+ \rightarrow 0^+_1) > 8.9~10^{22}$~yr and $\mathcal{T}_{1/2}^{0\nu}(0^+ \rightarrow 2^+_1) > 1.6~10^{23}$~yr, right-handed currents (V+A): $\mathcal{T}_{1/2}^{0\nu} > 5.4~10^{23}$~yr and Majoron emission: $\mathcal{T}_{1/2}^{0\nu} > 2.7~10^{22}$~yr.

\end{document}